\documentstyle[preprint,prd,aps]{revtex}
\begin{document}
\title{\bf Non-Perturbative Renormalization Group Flows in\\
           Two-Dimensional Quantum Gravity }
\author{\bf Ray L. Renken }
\address{Department of Physics, University of Central Florida,
Orlando, Florida, 32816 }
\author{\bf Simon M. Catterall}
\address{Theory Division, CERN CH-1211, Geneva 23, Switzerland}
\author{\bf John B. Kogut}
\address{Department of Physics, University of Illinois,
Urbana, Illinois 61821}
\date{\today}
\maketitle
\mediumtext
\widetext
\begin{abstract}
\begin{quotation}

Recently a block spin renormalization group approach was proposed for the
dynamical triangulation formulation of two-dimensional quantum gravity.
We use this approach to examine non-perturbatively a particular
class of higher derivative actions for pure gravity.

\end{quotation}
\vskip .25in
\end{abstract}
\newpage
\section{Introduction}
Dynamically triangulated random surfaces provide a lattice representation of
two-dimensional quantum gravity \cite{bkkm,dav}. Both in
the continuum and on a simplicial lattice
the usual Einstein action based on the Ricci scalar
is a topological invariant. 
Thus the simplest action for the lattice
theory at fixed volume and genus can then
be taken as zero. 

In principle, it is possible to add other
operators to this lattice action which are consistent
with the underlying symmetries of the model - here reparametrization invariance.
The lattice action would then take the form $S=\sum_i \beta_i O_i$ where
$\{O_i\}$ are a set of generic operators with associated coupling
constants $\{\beta_i\}$. For example, it is natural to consider
operators which are the lattice analogues of higher derivative terms -- 
integrals of powers of the scalar curvature.
In general these actions may then possess one or more
critical points $\{\beta_i^c\}$
in the coupling constant space where it may be possible to
construct continuum limits for the model.

The usual theory of two dimensional quantum gravity is
constructed about the special point $\beta_i=0$. Perturbation
theory then indicates that the higher operators are all
irrelevant in the renormalization group sense -- that is the long distance
continuum physics of models with $\beta_i$ non-zero is identical
to that at the fixed point $\beta_i=0$. Unfortunately, perturbation
theory can tell us nothing, in principle, about the
existence and properties of other fixed points situated
in regions of the parameter space where any of the $\beta_i$ are not small.
To probe such regions a nonperturbative procedure is required.
For conventional statistical mechanical models the block
spin renormalization group is one such technique \cite{wilson}. 
In this technique, a local kernel is used to construct an effective theory
with a carefully controlled change of scale which allows the calculation of
critical couplings and critical exponents.

Such a block spin formalism has recently been developed for
dynamical triangulations and applied to two-dimensional quantum
gravity coupled to Ising spins \cite{rr}. In contrast, a heuristic
renormalization group inspired approach has been advocated in \cite{kryz}.
In this paper, we apply the
block spin
renormalization group approach to pure quantum gravity. The
aim is to explore
the fixed point structure of the lattice model when a particular
class of higher derivative operator is included in the action.
Specifically, we use an action
$S=\alpha\sum_i \ln q_i$ where $q_i$ is the coordination number of
a site. Such a term consists of an infinite series of powers of
the curvature and arises naturally when we couple the theory to
scalar fields. We show that
the approach does indeed yield an appropriate fixed point and present
results which give strong evidence for the {\it nonperturbative}
irrelevance of such higher order curvature terms.

\section{Block Spin Renormalization Group}

The details of the algorithm are given in \cite{rr}. Here, we just
summarize the approach. The traditional way of implementing the
renormalization group within a numerical simulation is to generate a
sequence of lattice field configurations which are distributed according
to the usual Boltzmann weight. Each of these is then progressively
coarsened in some way which preserves the long distance physics.
Corresponding to each initial fine lattice configuration a succession
of `blocked' lattices is thus generated. Typically, the fields on
each `blocked' lattice are determined by the fields of the
lattice at one less blocking level. By examining the flows of
expectation values of a set of operators and their correlators
as a function of blocking level, it is then possible to
extract the critical couplings and critical exponents.

The choice of an apt `blocking' transformation is a very
important issue. For the case of random triangulations, the
lattice itself is the dynamical object. We thus require 
an algorithm for replacing a given random mesh with a succession
of coarsened descendents with approximately the same long distance
features. The most natural way to measure distance in this
context is by defining all lattice links to have length unity.
The distance between any two points is then taken as the
geodesic length between them -- the length (in lattice units)
of the shortest path connecting them on the lattice.

In order for the blocking algorithm to be apt it must be able to
replace a given mesh by one with a subset of the nodes triangulated
in such a way that the {\it relative lengths of blocked
geodesics reflect the underlying geodesic structure}. That is, like
a metric, the blocked triangulation tells us which points are
near and which are far apart and this must
accurately reflect the situation on the underlying
lattice. It appears to be a hard problem to give a rule which
when applied to an arbitrary random lattice accomplishes
this task. Our method, however, relies on a simple, local, iterative 
procedure
to generate the coarsened lattices.

Suppose, by some method, it has been possible to generate
a set of blockings of a given triangulation. In order to generate
a Monte Carlo sample, the fine lattice (blocking level zero) is then
updated using the stochastic link flip algorithm. In order that
the coarsened lattices reflect the new fine lattice it is necessary
to perform block link flips according to some suitable rule. This
rule then ensures that they `follow' the parent lattice as it
is updated.
Denote
a generic lattice at blocking level $k$ by $T_k$ and its successor
at level $k+1$ by $T_{k+1}$. 
Thus, any rule which specifies when to flip links in $T_k$ in response to flips
of the links in $T_{k-1}$ provides a definition of the
blocking transformation. An apt rule appears to
be to flip a block link in $T_k$ whenever that would connect two points 
that are closer
(on the lattice $T_{k-1}$) than the two currently linked. This
process is iterated recursively to generate a tower of
blocked lattices for each fine lattice.
This block rule ensures
that a given block lattice is determined from its `parent' at one
less blocking level in such a way that the relative
distance of blocked nodes is preserved.

There are two convenient
ways to choose the original lattice and its blocked form.  One is to start
with a regular lattice and to choose distinct subsets of points 
(those corresponding to a usual square lattice
blocking) that can obviously
be triangulated in a regular way.  The other is to start with a triangulation
that is viewed as the block lattice and to add as many points as desired to
produce the bare lattice.  Updating the block lattice with a number 
of block link sweeps then
relaxes the block lattice.  

The Monte Carlo cycle thus begins with an update sweep of the
fine lattice followed by a number of applications of the block link
update rule (typically five to ten block link sweeps) at each
blocking level.

Any expectation values computed on a blocked lattice can be viewed as coming
from an effective action.  There is a sequence of effective actions that
corresponds to the sequence of blocking levels.  If the original action is
critical (and if the renormalization group transformation is apt), this
sequence converges to a fixed point.  Such should be the case for dynamical
triangulations with action equal to zero.  Such should also be the case if
any irrelevant term is added to the action.  In this case, the sequence should
converge, not just to any fixed point, but to the same fixed point obtained
without the irrelevant terms.

In practice, although the effective actions converge to a fixed point when
the theory is critical, the expectation values obtained on the block lattices
do not.  This is because each renormalization group transformation reduces
the size of the lattice and hence increases the finite size effects.  A single
sequence of blocking levels with their corresponding expectation values will
not display convergence toward a fixed point.  However, two sequences,
beginning with bare lattices of different volumes can do this.  The trick is
to choose the bare lattice of one of the sequences to have the same volume as
the first blocked level of the other sequence.  In this way, expectation
values can be compared on lattices with the same volume (and therefore the
same finite size effects) but with different numbers of iterations of the
renormalization group transformation.  Since the finite size effects are
identical, any difference in expectation values can only be due to a difference
in effective actions.  As the renormalization group transformation is
iterated and the actions flow toward a fixed point, the difference in
effective actions should rapidly decrease yielding a progressively smaller
difference in 
expectation values.

\section{Results}

Our first goal, then, is to implement the block spin renormalization group
transformation described above on dynamical triangulations with an action
equal to zero and to see if the matching procedure just outlined produces
pairs of expectation values that are increasingly close as the blocking
level is increased.  Seven operators are used in this study.  The first
six are all powers or correlations of the coordination number at a site
($q_i$) minus six (its flat-space, regular lattice value) and are all
normalized by the number of links.  The first is the nearest neighbor
correlation:
$$O_1 = \sum_{<ij>} (q_i - 6)(q_j - 6).$$
The second is the correlation between the nodes conjugate to
a link (the nodes that the link would join if it were flipped):
$$O_2 = \sum_{<ij>} (q_{i^{\prime}} - 6)(q_{j^{\prime}} - 6)$$
(where $i^{\prime}$ and $j^{\prime}$ represent the nodes where the ends of the
flipped link would go).  The third is the product of the first two:
$$O_3 = \sum_{<ij>} 
        (q_i - 6)(q_j - 6)(q_{i^{\prime}} - 6)(q_{j^{\prime}} - 6).$$
The fourth, fifth, and sixth are the second, third, and fourth powers of the
coordination number minus six:
$$O_4 = \sum_{i} (q_i - 6)^2, \quad O_5 = \sum_{i} (q_i - 6)^3,\quad 
O_6 = \sum_{i} (q_i - 6)^4.$$
Finally, the seventh operator is the maximum coordination number of the
lattice:  
$$O_7 = {\rm max}(q_i)$$
Lattices were used with 9, 36, 144, 576, and 2304 nodes which 
allowed for up to four iterations of the blocking transformation.  
The results shown correspond to $1\times 10^5$ bare lattice
sweeps. Table 1
shows the expectation values at all blocking levels starting from the largest
lattice.  There is a great variation in the expectation values as a function
of block level and it is not at all obvious that they are approaching a fixed
point.  The matching can be seen in table 2 which
compares the seven expectation values after three and four iterations of
the blocking transformation on lattices such that the final number of nodes
is nine.  They match fairly well, an indication that the effective theory
is near its fixed point.  Figure 1 uses expectation value differences of
$O_7$ to give a graphical representation of the approach to the fixed point.

Now consider a perturbation of this scenario using the action
$$S = \alpha \sum_i \ln(q_i) $$
If this term is irrelevant, the sequence of expectation values generated
by iterating the blocking transformation should approach those generated
with $S = 0$ at large blocking levels, even if the expectation values differ
a great deal at the lower blocking levels.  Table 3 shows the data in the
case of $\alpha = -1$.  Figure 2, using $O_7$ again, gives a graphical
representation of this data along with data for $\alpha = +1$.  The fact
that the expectation value differences approach zero as the blocking level
increases confirms that $\ln(q)$ is indeed an irrelevant operator.  The
results are similar for much larger $\alpha$.  Figure 3 shows the analogous
data for $\alpha = \pm 10$.  At a true fixed point, all of the expectation
values should match, not just one.  Figures 4 and 5 give the expectation
value differences for $O_1$ at the same values of $\alpha$ as in figures 2
and 3 respectively.  Matching is demonstrated for this operator as well.

In \cite{bkkm} there is evidence that for negative enough $\alpha$ there is
a phase transition to some crumpled state.  Such a transition is not visible
in perturbation theory \cite{kawai}. If such a transition exists,
one would expect the expectation values to flow to a set of values different
from those obtained with $S = 0$.  We find that while at negative values of
$\alpha$ the expectation values on the bare lattice start looking dramatically
different from those at $S=0$ (for instance the value of $O_7$ increases by
more than an order of magnitude) the renormalization group trajectories flow
to the same point within statistics.  

Thus, the renormalization group scheme
used here gives no evidence for a phase transition.  It is possible that there
is such a transition and that either the particular renormalization group
transformation used here is not ``apt'' for that transition or the expectation
value differences of the blocked operators are smaller than our errors.  The
statistical uncertainty of the
$\alpha = -10$ data at the highest level of blocking is from three to five
times larger (depending on the operator) than that of any of the other values
of $\alpha$ considered in this paper.  In this regard, it should be noted that
the effects of Ising matter at the critical point on expectation values in the
gravitational sector are too small to be detected with current statistics.
However, the effects of matter on the gravitational sector are notoriously
small for this formulation of quantum gravity whereas the higher derivative
term can clearly have a strong effect.  It may be that there is a transition
that is nearby in the space of theories possibly of a higher
order of multicriticality. To see such a fixed point would require
tuning of additional couplings.

To summarize, we have presented results concerning the fixed point 
structure of the dynamical triangulation model for two dimensional
quantum gravity. These have been obtained using an adaptation of the 
Monte Carlo renormalization group to the situation where the lattice
itself carries the dynamical degrees of freedom. 
We have studied a class of higher derivative operator and
given evidence that such an operator is truly irrelevant outside of
perturbation theory. We see no evidence for new fixed points or
equivalently new phase transitions in the lattice model. It is possible
however, that other choices of higher derivative operator might indeed show
new structure \cite{sumit}.  The technique used in this paper can easily be
applied to other actions as well.

\acknowledgements
We thank Sumit Das for helpful discussions.  We acknowledge use of computer
time on the Cray YMP located at Florida State University.

\vskip .5in
\centerline{List of Figures}
\vskip .5in
\begin{enumerate}
\item The difference between expectation values of the maximum coordination
number, $O_7$, computed on lattices of the same size but for systems that
differ (by one) in the number of times they have been blocked.  The blocking
level listed is that of the system that has been blocked the most.  The
original action is zero.

\item The difference between expectation values of the maximum coordination
number, $O_7$, computed with two different actions as a function of the
blocking level.  The diamonds represent expectation values obtained with
$S = \alpha \sum_i \ln(q_i)$ minus those obtained with $S = 0$ when
$\alpha = +1$ while the squares represent the analogous results for
$\alpha = -1$.

\item This figure is like figure two except that the squares represent
$\alpha = +10$ and the crosses represent $\alpha = -10$.  Crosses are missing
for levels zero and one because the data is off scale by more than an order
of magnitude.

\item This is the same plot as figure 2 except that $O_1$ is used instead
of $O_7$.

\item This is the same plot as figure 3 except that $O_1$ is used instead
of $O_7$.  Again, some of the $\alpha = -10$ data is off scale.

\end{enumerate}
\vskip .5in
\mediumtext
\begin{table}
\begin{tabular} {cddddd} 
operator & $n = 0$&  $n = 1$   &  $n = 2$   &  $n = 3$   &  $n=4$   \\ \hline
$O_1$ &  1.499(1) &  4.99(5)   &  4.33(6)   &  1.12(5)   & -0.367(7)\\
$O_2$ &  2.611(2) &  6.68(5)   &  8.3(2)    &  4.2(1)    &  0.35(2) \\
$O_3$ & -4.63(2)  &  10.9(5)   &  -2(1)     &  0(1)      &  0.19(8) \\
$O_4$ &  3.500(1) &  5.01(2)   &  5.42(8)   &  3.91(6)   &  0.70(1) \\
$O_5$ &  22.25(2) &  56.5(8)   &  70(3)     &  26(1)     & -0.58(3) \\
$O_6$ &  328.7(7) &  1400(30)  &  1900(200) &  370(20)   &  3.8(1)  \\
$O_7$ &  29.87(3) &  35.2(2)   &  28.8(4)   &  17.4(2)   &  7.89(1) \\
\end{tabular}
\caption{ Expectation values of seven operators at all blocking levels
beginning with a $2304$ node lattice. The action is zero. }
\end{table}
\narrowtext
\begin{table}
\begin{tabular} {cdd} 
operator & $V = 2304$ & $V = 576$ \\ \hline
$O_1$    & -0.367(7) & -0.363(4)  \\
$O_2$    &  0.35(2)  &  0.34(1)   \\
$O_3$    &  0.19(8)  &  0.27(4)   \\
$O_4$    &  0.70(1)  &  0.677(5)  \\
$O_5$    & -0.58(3)  & -0.52(1)   \\
$O_6$    &  3.8(1)   &  3.54(5)   \\
$O_7$    &  7.89(1)  &  7.881(6)  \\
\end{tabular}
\caption{ The expectation value of the seven operators on two lattices which
have been blocked three and four times, respectively.  The original volumes
were chosen so that the final blocked lattices have the same number of nodes
(nine). }
\end{table}
\mediumtext
\begin{table}
\begin{tabular} {cddddd} 
operator & $n = 0$&  $n = 1$   &  $n = 2$   &  $n = 3$   &  $n=4$    \\ \hline
$O_1$ &  2.295(2) &  5.10(6)   &  4.43(6)   &  1.12(5)   & -0.367(8) \\
$O_2$ &  4.138(4) &  7.16(8)   &  8.4(2)    &  4.5(2)    &  0.34(2)  \\
$O_3$ & -10.93(5) &  11.1(6)   &  -2(2)     &  0(1)      &  0.31(8)  \\
$O_4$ &  4.360(2) &  5.16(3)   &  5.47(5)   &  4.04(8)   &  0.694(8) \\
$O_5$ &  38.68(7) &  66(1)     &  71(2)     &  28(1)     & -0.56(2)  \\
$O_6$ &  727(3)   &  2000(70)  &  1900(100) &  416(30)   &  3.71(7)  \\
$O_7$ &  36.79(5) &  37.5(3)   &  29.1(2)   &  17.9(2)   &  7.89(1)  \\
\end{tabular}
\caption{ Expectation values of seven operators at all blocking levels
beginning with a $2304$ node lattice. The action is $S = -\sum_i \ln(q_i)$ }
\end{table}

\begin{references}

\bibitem{bkkm} D. V. Boulatov, V. A. Kazakov, I. K. Kostov, and A. A. Migdal,
               Nucl. Phys. B275, 641, 1986.
\bibitem{dav}  F. David, Nucl. Phys. B257, 45, 1985.
\bibitem{wilson} K. G. Wilson, Rev. Mod. Phys. 47, 773 (1975).
\bibitem{rr} Ray L. Renken, ``The Block Spin Renormalization Group Approach
             and Two-Dimensional Quantum Gravity", hep-lat/9405007, Phys. Rev.
	     D in press.
\bibitem{kryz} D. A. Johnston, J. P. Kownacki and A. Krzywicki ``Random
               Geometries and Real Space Renormalization Group '', 
	       hep-lat/9407018.
\bibitem{kawai} H. Kawai and R. Nakayama, ``Quantum $R^2$ gravity in two
                dimensions'', KEK-TH-355 1993.
\bibitem{sumit} S. Das et al. Mod. Phys. Lett A5, 1041, 1990.
\end{references}
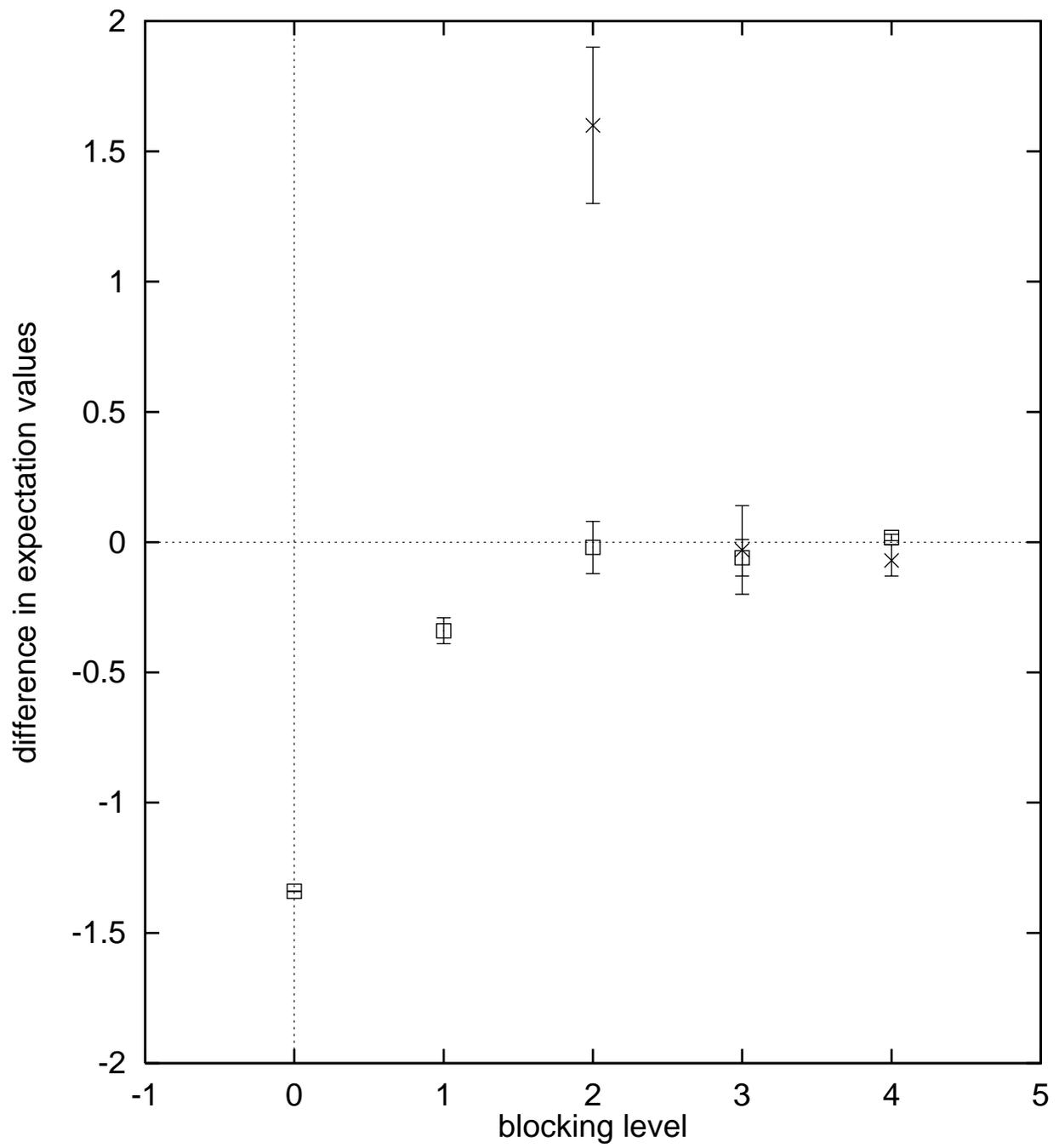
\end{document}